\documentclass[final,nofootinbib,superscriptaddress]{revtex4-1} 
\usepackage{graphicx,amsmath}
\usepackage{relsize}
\usepackage{import}
\usepackage{color}
\usepackage{listings}
\newcommand{\sgn}{\operatorname{sgn}}

\begin{document}
\title{Relativistic Hydrodynamics on Graphic Cards}
\author{Jochen Gerhard}
\affiliation{Frankfurt Institute for Advanced Studies,
  Ruth-Moufang-Stra\ss{}e 1, 60438 Frankfurt am Main}
\affiliation{Institut f\"ur Informatik, Johann Wolfgang Goethe-Universit\"at, Robert-Mayer-Stra\ss{}e 11--15, 60054 Frankfurt am Main}
\email{jochen.gerhard@compeng.uni-frankfurt.de}
\thanks{Tel.: +4969 798 44112}
\thanks{Fax: +4969 798 44109}
\author{Volker Lindenstruth}
\affiliation{Frankfurt Institute for Advanced Studies, Ruth-Moufang-Stra\ss{}e 1, 60438 Frankfurt am Main}
\affiliation{Institut f\"ur Informatik, Johann Wolfgang Goethe-Universit\"at, Robert-Mayer-Stra\ss{}e 11--15, 60054 Frankfurt am Main}
\author{Marcus Bleicher}
\affiliation{Frankfurt Institute for Advanced Studies, Ruth-Moufang-Stra\ss{}e 1, 60438 Frankfurt am Main}
\affiliation{Institut f\"ur Theoretische Physik, Johann Wolfgang Goethe-Universit\"at, Max-von-Laue-Stra\ss{}e 1, 60438 Frankfurt am Main}

\begin{abstract}

We show how to accelerate relativistic hydrodynamics simulations
using graphic cards (graphic processing units,
\textsmaller{GPUs}). These improvements are of highest relevance e.g. to the field of high-energetic nucleus-nucleus collisions
at \textsmaller{RHIC} and \textsmaller{LHC} where (ideal and dissipative) relativistic
hydrodynamics is used to calculate the evolution of hot and dense
\textsmaller{QCD} matter. The results reported here are based on the Sharp
And Smooth Transport Algorithm \textsmaller{(SHASTA)}, which is employed in many hydrodynamical
models and hybrid simulation packages, e.g. the Ultrarelativistic Quantum
Molecular Dynamics model
\textsmaller{(UrQMD)}. 
We have redesigned the \textsmaller{SHASTA} using the
\textsmaller{OpenCL} computing framework to
work on accelerators like graphic processing units \textsmaller{(GPUs)}
as well as on multi-core processors. With the redesign of the algorithm the
hydrodynamic calculations have been accelerated by a factor 160
allowing for event-by-event calculations
and better statistics in hybrid calculations.
\end{abstract}
\maketitle
\section{Introduction}
Relativistic hydrodynamics\cite{springerlink:10.1023/A:1021151905577} has a long history since its first
application in nuclear collisions\cite{springerlink:10.1007/BF02745507} and for the evolution
of the universe\cite{1979sgrr.work..423W}.
It is still an excellent tool to describe systems of different scales, ranging from
collisions of galaxies down to collisions of heavy ions or even
protons at \textsmaller{LHC}\cite{PhysRevC.83.044915}.
All these simulations need to respect the relativistic nature of the
dynamics, however the focus lies on different aspects. While astronomical
calculations have to include viscosity and turbulences, heavy ion simulations
can neglect turbulences. Here, algorithms that have the ability to
capture shock phenomena or relatively sharp gradients (and
potentially very small viscosity) have to be applied. For the present
paper we focus on nuclear collisions, however, the methods can also be
transfered to cosmological and supernova simulations.

A large body of current numerical tools to explore and interprete the experimental
data obtained in high energy experiments employs relativistic
hydrodynamics in various facets: On the purely hydrodynamic side there are the
models of Heinz and Kolb\cite{Kolb:2003dz},
the model of Romatschke\cite{Baier:2006um},  Csernai's\cite{Csernai:2011gg}
hydrodynamic model, and various multi fluid approaches,
e.g. by Toneev\cite{Ivanov:2005yw, Toneev:2008pq}, 
and Brachmann and Dumitru\cite{Dumitru:2000ai}.
In the recent years also hybrid approaches between Monte-Carlo /
Boltzmann event simulators and hydrodynamics have gained substantial
popularity. Most noteworthy in this respect are the approaches of Hirano
and Nara (hydrodynamics+\textsmaller{JAM})\cite{Hirano:2012kj},
Bleicher and Petersen \textsmaller{(UrQMD 3.3)}\cite{Petersen} using
the \textsmaller{SHASTA}, 
\textsmaller{NeXSPheRIO}\cite{ANDRADE2007} based on smoothed particle hydrodynamics,
\textsmaller{EPoS}\cite{Werner:2010aa}, Bass and Nonaka\cite{Nonaka:2005aj} realizing a
Lagrange approach coupled to \textsmaller{UrQMD},
the hydro-kinetic model \textsmaller{(HKM)}\cite{Karpenko201050},
and \textsmaller{MUSIC}\cite{Schenke:2010nt}.
For the numerical solution different methods are applied ranging from
smooth particle methods, over the Kurganov-Tadmor\cite{Kurganov2000241} algorithm, to the
\textsmaller{SHASTA}\cite{shasta}. The present examples are based on the
\textsmaller{SHASTA} as it is employed in \textsmaller{UrQMD},
which offers since version \textsmaller{UrQMD 3.3} a hybrid transport
model\cite{Petersen}, based on the relativistic implementation by
Rischke\cite{Rischke}. We shall refer to this implementation by the
name of classical \textsmaller{FORTRAN} implementation, as it is coded
completely in \textsmaller{FORTRAN 77} compared to our
\textsmaller{C++ / OpenCL} implementation.

Although the reduction of complexity by neglecting turbulences and
viscosity reduces the computational demands of the simulation, the
running time, typically between 20 minutes and one hour per event for a $(3+1)$ dimensional
simulation, still prohibits the calculation of high statistics samples.
As the \textsmaller{SHASTA} is well tested and suitable for heavy ion
collisions, we have redesigned this algorithm to run on
modern accelerator hardware in order to improve the performance. The presented
\textsmaller{OpenCL} implementation allows us to harvest the computational
power of accelerators, \textsmaller{GPUs}, and multi-core
processors. After all improvements described in this paper, we
gain a speed-up of up to a factor 160 for the full
propagation on a \textsmaller{GPU}, compared to the standard
\textsmaller{FORTRAN} implementation used in \textsmaller{UrQMD 3.3} 
on a \textsmaller{CPU}.
For previous implementations of classical hydrodynamics algorithms on
\textsmaller{GPUs} we refer the reader to \cite{Kolb05dynamicparticle,
  citeulike, springerlink:10.1007/11758549_34, Frezzotti}.
\section{Formulation}
The hydro phase of heavy ion collisions is governed in principal by
the Euler equations:
\begin{align}
\label{eq:emass}
\frac{\partial \rho}{\partial t} + \nabla \cdot \left(\rho \mathbf{v}
\right) &= 0\\
\label{eq:emoment}
\frac{\partial\rho\mathbf{v}}{\partial t} + \nabla \cdot \left(
  \mathbf{v} \otimes \left(\rho\mathbf{v}\right)\right) + \nabla p &=
0\ \\
\label{eq:eenergy}
\frac{\partial \varepsilon}{\partial t} + \nabla
\cdot\left(\mathbf{v}\left(\varepsilon+p\right)\right) &= 0
\end{align}
With $\rho$ being the density, $\mathbf{v}$ the velocity vector, $p$
the pressure, and $\varepsilon$ the energy density.
The equations are closed by adding an equation of state. For the heavy
ion collision case different equations of state have been proposed and
can be used for the simulation. Different models use non analytic
equations of state, and provide parametrized equations for the
thermodynamical quantities. Various numerical interpolations schemes
can be applied here and leave the possibility for additional optimizations
on \textsmaller{GPUs}. For the present numerical study we
restrict ourselves to the ideal gas equation of state.

As high energy heavy ion collisions proceed near to the speed of light, relativistic
effects have to be taken into account, which reformulates the
equations to:
\begin{align}
 \partial_\mu T^{\mu\nu} &= 0 \\
  \partial_\mu J_B^\mu &= 0
\end{align}
For the propagation of the energy-momentum tensor $T^{\mu\nu}$ and the
baryon current $J_B^\mu$.
The energy-momentum-tensor can be written
as:
\begin{align}
T^{i j} =& \left(\varepsilon + p\right)\gamma^2v_iv_j + p\delta_{i j} \\
T^{0 i}=& \left(\varepsilon + p\right)\gamma^2v_i\\
T^{0 0}=& (\varepsilon + pv^2)\gamma^2 
\end{align}
And taking $J_B^{\mu} = \left(\gamma\rho, \gamma\rho\mathbf{v}\right)
= \left(\mathcal{N},\mathcal{N}\mathbf{v}\right)$ the
energy-momentum-tensor and the Baryon density can be written in terms
like mass, momentum and energy:
\begin{align}
\label{eq:cf-density}
\mathcal{N} =& \rho\gamma\\
\label{eq:cf-momentum}
\boldsymbol{\mathcal{M}} =& \left(\varepsilon + p\right)\gamma^2\mathbf{v}\\
\label{eq:cf-energy}
\mathcal{E} =& (\varepsilon + pv^2)\gamma^2
\end{align}
Hence, following \cite{csernai1994introduction, Ollitrault:2007du}, the relativistic Euler equations take the form similar to the non
relativistic differential equations \eqref{eq:emass},
\eqref{eq:emoment}, and \eqref{eq:eenergy}:
\begin{align}
\label{eq:relmass}
\frac{\partial}{\partial t} \mathcal{N} + \nabla\cdot
\left(\mathcal{N}\mathbf{v}\right) &= 0 \\
\label{eq:relmoment}
\frac{\partial}{\partial t} \boldsymbol{\mathcal{M}} +
\nabla\cdot\left(\boldsymbol{\mathcal{M}}\mathbf{v}+p\mathbf{\overline{\overline{I}}}\right)
&= 0 \\
\label{eq:relenergy}
\frac{\partial}{\partial t} \mathcal{E} +
\nabla\cdot\left(\mathbf{v}\left(\mathcal{E} + p\right)\right) &= 0
\end{align}
\section{Numerical Method}
The relativistic \textsmaller{SHASTA}\cite{Petersen, Rischke} is composed of four phases:
\begin{enumerate}
  \item Geometric transport.
  \item Anti-Diffusion with flux limiter.
  \item Relativistic correction.
  \item Relativistic calculation of the equation of state.
\end{enumerate}
The geometric transport resembles a finite volume scheme by
calculating fluxes between cell boundaries. For an arbitrary quantity
$U$ (standing for $\mathcal{E}$, $\boldsymbol{\mathcal{M}}$, and
$\mathcal{N}$ of the
equations~\eqref{eq:relmass},~\eqref{eq:relmoment}), and~\eqref{eq:relenergy})
the fluxes to neighboring cells are calculated by geometrical
factors. These factors are determined by the pressure and velocity of
Lagrangian fluid parcells\cite{shasta}. After the propagation step, the
parcells are interpolated back an equidistant lattice with space
index $j$ and time index $n$.
The geometrical factors including the interpolation back to the grid
are given by\cite{Rischke}:
\begin{align}
\label{eq:shasta-diff}
Q_{\pm} &= {\frac{\frac{1}{2} \mp v^{n+\frac{1}{2}}_j \cdot \lambda}{1
  \pm ( v^{n+\frac{1}{2}}_{j \pm 1} \cdot \lambda - v^{n+\frac{1}{2}}_j \cdot \lambda)}} .
\end{align}
With $\lambda = \frac{\Delta t}{\Delta x}$ being the
Courant-Friedrichs-Lewy number, and $v$ the
propagation velocity, calculated on a staggered grid in time direction. 

The propagation of energy and momenta consists not only
of the material derivative, but an additional source term $f$
originating from the acceleration of fluid particles by the pressure
gradient. Thus, for the propagation of momenta
(equation~\eqref{eq:relmoment}) the additional divergence of the
pressure tensor $p\mathbf{\overline{\overline{I}}}$
and for the propagation of energy (equation~\eqref{eq:relenergy}) the
divergence of  velocity and pressure $\mathbf{v}p$ have to be
calculated in order to compute a time step.
The source term's differential $\Delta(f)$ is also computed on a staggered
grid in time direction. We use the central differential $\Delta(f) =
-\frac{1}{2}(f^{n+\frac{1}{2}}_{j+1}-f^{n+\frac{1}{2}}_{j-1})$ as
proposed in \cite{Rischke}. 
The purely geometrically propagated quantity $\hat U_j^n$ takes the following form:
\begin{align}
\label{eq:shasta-geo}
\Delta_j &= U^n_{j+1} - U^n_j , \\
\hat U_j^n &=
\frac{1}{2}Q^2_+\Delta_j-\frac{1}{2}Q^2_-\Delta_{j-1}+(Q_++Q_-)U_j^n+\lambda\Delta(f) .
\end{align}
As the geometric transport produces inherently a certain amount of
numeric diffusion the next step is to correct the results by an
anti-diffusion term\footnote{Different anti-diffusion terms are
  possible. As stated in \cite{Rischke} the most suitable for this
  case is the so called \emph{phoenical} anti-diffusion.}
$A^{\text{ph}}$. The constraint to the anti-diffusion is not to create
new maxima or minima on the grid. Thus, the anti-diffusion itself is limited by a flux limiter depending
on a neighborhood of points. With the preceding definitions, a time
step is completely determined by:
\begin{align}
\hat \Delta_j &= \hat U_{j+1} -\hat U_j\\
A^\text{ph}_j &= \frac{1}{8}\left(\hat\Delta_j -
  \frac{1}{8}\left(\Delta_{j+1}-2\Delta_j+\Delta_{j-1}\right)\right)\\
\sigma &= \sgn A^{\text{ph}}_j\\
A_j &= \sigma \cdot \max\left\{0,
  \min\left\{\sigma\hat\Delta_{j+1},|A_j^{\text{ph}}|,\sigma\hat\Delta_{j-1}\right\}\right\}
\\
U_j^{n+1} &= \hat U_j^n - A_j + A_{j-1}
\label{eq:shasta-n}
\end{align}
In the relativistic case, all quantities are boosted from
computational frame to their eigenframe in order to calculate the thermodynamic
pressure, the baryo-chemical potential, and the propagation velocity. To
obtain these quantities from the energy density $\varepsilon$ and  
the baryon density $\rho$ we employ the ideal gas equation of state. Although the ideal
gas equation enables an analytical solution to the calculation of the
propagation velocity, we have used a fixed point root finding
algorithm to allow for an easy implementation of a tabled equation of
state without major changes in the codebase.
\subsection{Algorithm Design}
The usage of \textsmaller{GPUs} had been of great interest to the
field of high energy physics even before state-of-the-art programming frameworks,
like \textsmaller{CUDA} or \textsmaller{OpenCL}, have been developed\cite{fodor}. 
In order to harvest best performance of \textsmaller{GPUs} one has to
bear in mind certain architectural constraints of these devices. This
implies often a very different approach than classical
\textsmaller{CPU} programming. The implementation in this paper is
designed in \textsmaller{OpenCL} and fitted to an \textsmaller{AMD}
5870 \textsmaller{GPU}. Therefore certain optimizations are also
different \cite{Daga:2011:AMO:2117686.2118481} to typical
optimizations carried out on \textsmaller{NVIDIA GPUs}.
Although designed for this special kind of \textsmaller{GPU} the
implementation still shows significant speed-up on
\textsmaller{CPU}-only systems. Here additional optimizations, respecting
caching and memory layout, e.g. interleaving the 3 dimensional
grid variables, would mitigate the losses due cache misses when
propagating in $y$ or $z$ direction, allow for further significant accelerations.

The \textsmaller{OpenCL-SHASTA}
consists of a \textsmaller{C++}-part, managing the memory allocation
and enqueueing of the \emph{kernels}. The kernels are routines
written in \textsmaller{OpenCL} and completely run on the
\textsmaller{GPU} or multi-core \textsmaller{CPU}. Kernels are
executed in a parallel manner, and each singular instance of a kernel
is called a \emph{work-item}. These work-items are mapped, in
hardware, to the stream cores of \textsmaller{GPUs} and
\textsmaller{CPUs}. The mapping occurs in small groups, whose size
depends on the hardware used. The smallest possible group is called 
\emph{wavefront}. Within each wavefront the execution flow must be
uniform, i.e. when branching occurs within a wavefront, all branches
are computed serially. 

The approach to parallelize on \textsmaller{GPUs} must bear in mind,
that the running program at the end consists not of a few parallel
threads, but merely thousands of concurrent execution streams. Yet most of
this streams are computing \textit{almost} the same, which ranks
this approach as a \emph{Single Instruction Multiple Data}
(\textsmaller{SIMD}) approach. In order to organize these thousands of
execution streams the first, and
arguably most important steps, are the \emph{problem decomposition}
and \emph{domain decomposition}.
Both points are somewhat entangled, still first order they can be
handled separately.

\subsubsection{Problem Decomposition}
\label{sec:problem}
\begin{figure}[ht]
\includegraphics[width=0.666\textwidth]{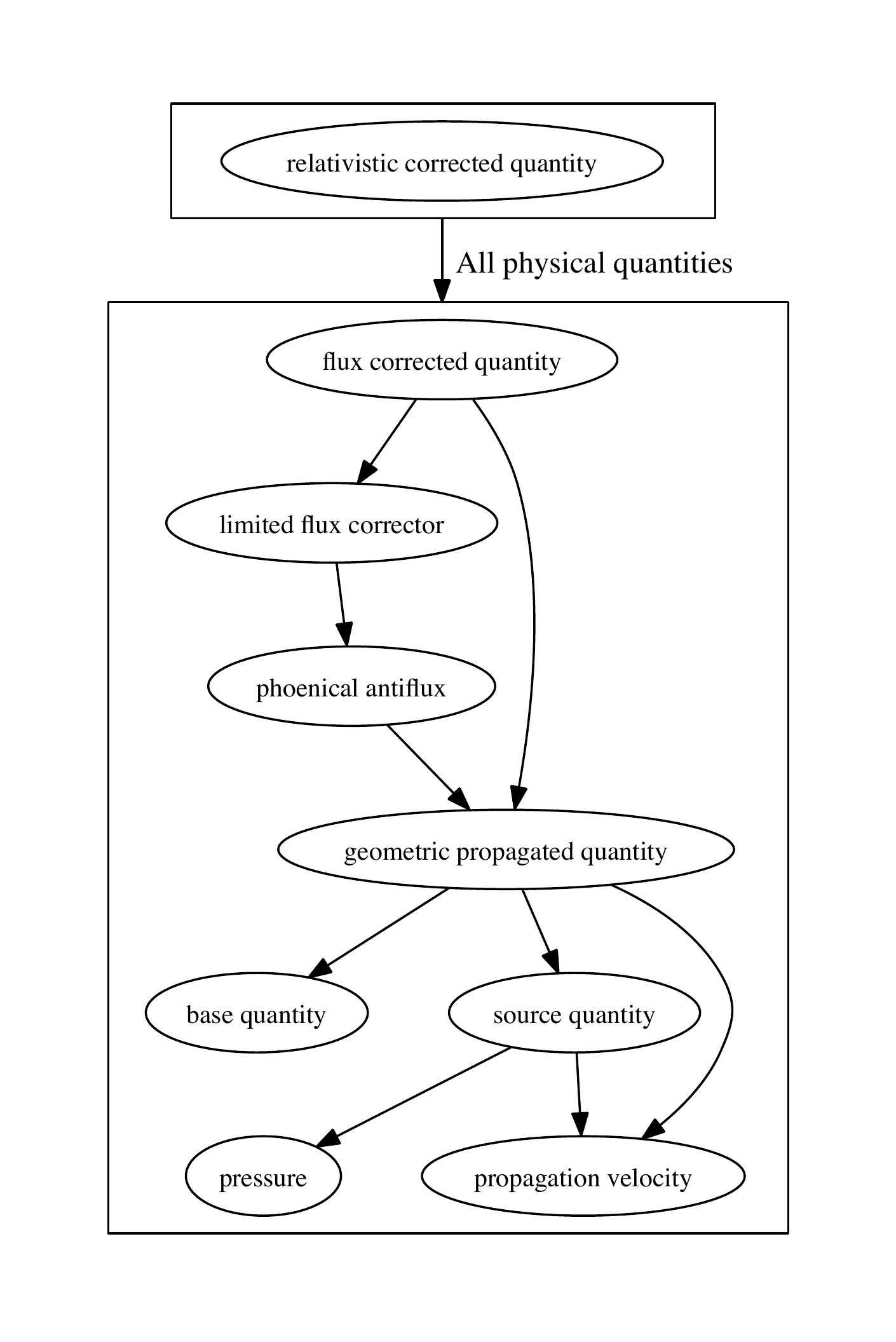}
\caption{\label{fig:flowgraph}: Variable dependencies in data flow, time
  and space indices are omitted. (Each arrow reads as ``depends on'')}
\end{figure}

Firstly the dataflow of the algorithm has to be analyzed. Not
only the physical quantities, that are finally stored, 
but also the steps in between and intermediate results of the
computation are important. In figure~\ref{fig:flowgraph} the
dependencies within each timestep are illustrated. Before each
quantity can be calculated, all the quantities pointed to, have to be
computed. The dataflow and the spacial
and causal dependencies of variables within build the
constraints to any possible parallel computation. 
The \textsmaller{SHASTA} performs the propagation of the
energy-momentum tensor and the net baryon current. Therefore five
physical quantities ($\mathcal{E}$, $\boldsymbol{\mathcal{M}}$, 
and $\mathcal{N}$) are propagated (equations~\eqref{eq:cf-density},
\eqref{eq:cf-momentum}, and \eqref{eq:cf-energy}). The full
propagation of each quantity is done by the sequent steps from
equation~\eqref{eq:shasta-diff} to
equation~\eqref{eq:shasta-n}. Subsequently relativistic corrections
have to be applied, as well as the calculation of the thermodynamic
quantities in their eigenframe.

\begin{figure}[t]
\includegraphics[width=0.666\textwidth]{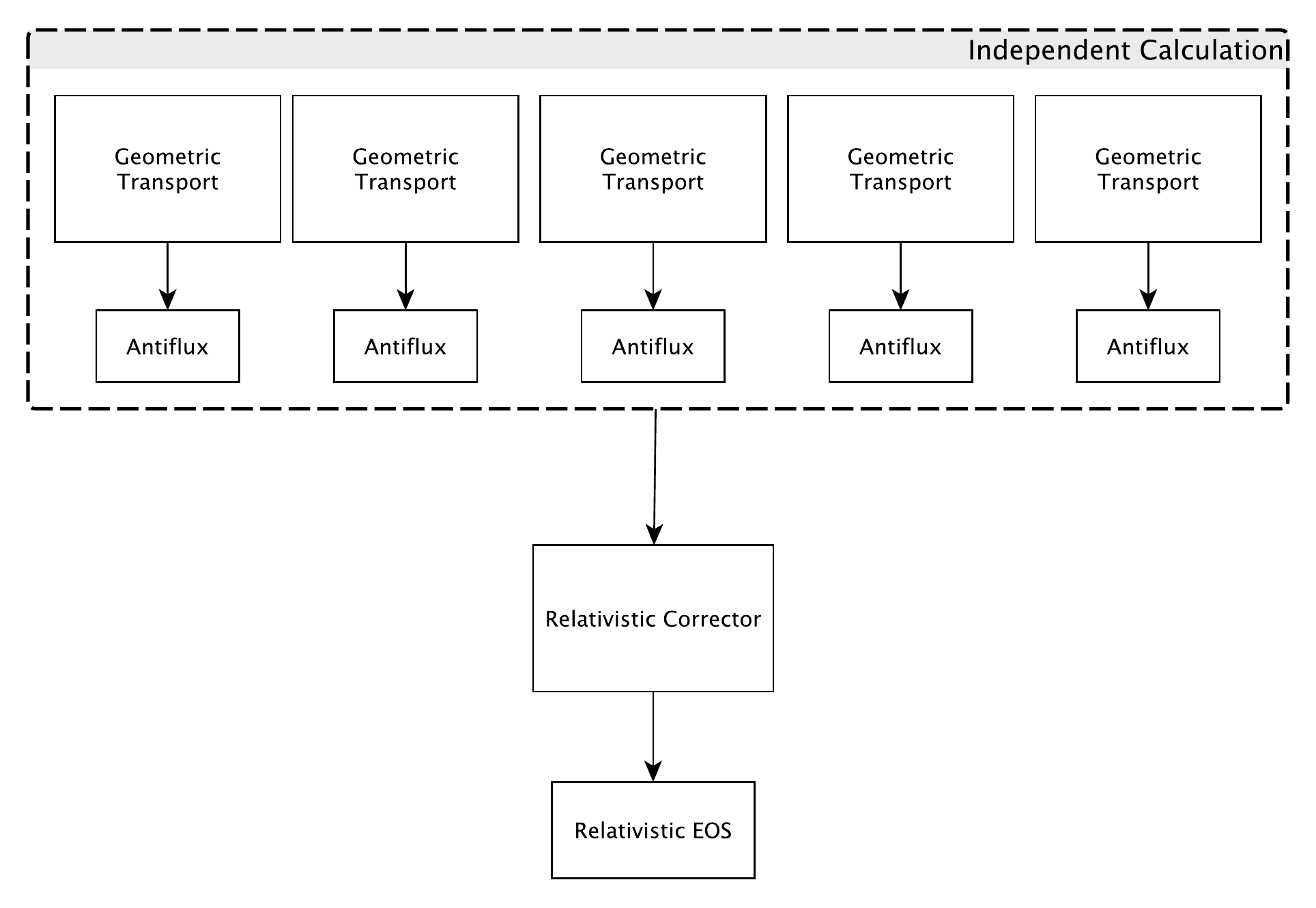}
\caption{\label{fig:flow}: Problem decomposition and data flow in \textsmaller{OpenCL-SHASTA}.}
\end{figure}

Most of the propagated quantities
can be calculated, completely in parallel. 
As illustrated in figure \ref{fig:flow}, the propagation of each
quantity, including the anti-diffusion and flux corrector, is 
calculated independently. Though, as the relativistic corrector needs
the propagated state of all quantities (see also
figure~\ref{fig:flowgraph}), the execution of the
relativistic corrector and the calculation of the equation of state is
scheduled after all five kernels, propagating the quantities
independently, have finished computation.
This is realized by the orchestrating method of the
\textsmaller{GPU}-class (listing \ref{enqueue}).

\label{alternative}
Another possible decomposition we have investigated on, is using the kernels firstly to
calculate all the geometric propagated quantities (equation~\eqref{eq:shasta-geo})
in parallel and subsequently applying new kernels for the anti-diffusion step.
Thereby one kernel can calculate more than one geometric propagated
quantity (e.g. all the momenta together) which favors the usage of the vector units of
\textsmaller{GPUs} and modern \textsmaller{CPUs}\footnote{Almost every
 modern processor offers a richt set of vector units
 (\textsmaller{SSE}).}. However this approach makes extensive usage of
the \textsmaller{GPU} memory, by storing intermediate results to the
global memory and reloading it in the anti-diffusion kernels
later. In the execution model of \textsmaller{OpenCL} a global
synchronization between work-units is not possible. Thus, different
kernels have to be enqueued to serialize the task.
\subsubsection{Domain Decomposition}
\label{sec:domain}
\begin{figure}[b]
\includegraphics[width=0.666\textwidth]{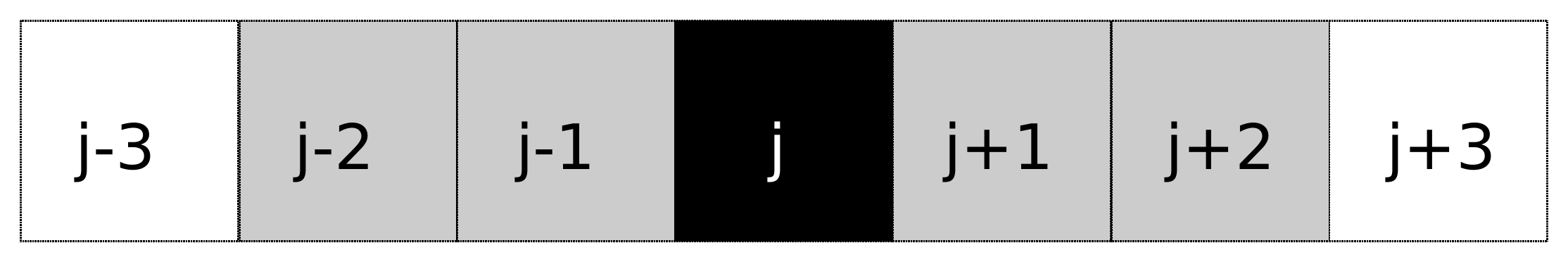}
\caption{The neighborhood used to compute the propagated quantity in
  cell~$j$ (black) spans seven different cells. To compute the flux corrected value for cell~$j$~the
  geometric transport for cells~$(j-2) \dots (j+2)$~(grey) is needed.}
\label{fig:neighbor}
\end{figure}
For every grid based algorithm the domain decomposition is often
suggested by the grid structure. Though the exact mapping of kernel
number to grid size depends on hardware and algorithm type.
The current implementation of \textsmaller{OpenCL-SHASTA} uses a
kernel per physical quantity for each of the eight million grid cells
cells. This choice works very well on \textsmaller{GPUs}. 
Optimizations aimed mainly on multi-core \textsmaller{CPUs}
would do more efficiently using kernels responsible for more than one
cell or more than one physical quantity per cell.
Depending on the algorithm type, see section~\ref{sec:problem}, the
dependencies within each grid cell might lead to even more kernels per
grid cell on \textsmaller{GPUs}.
The $(3+1)$-dimensional problem of relativistic hydrodynamics is
perfectly suitable to the three dimensional \emph{NDRange} arguments
in \textsmaller{OpenCL}. 
In order to implement the $(3+1)$-dimensional propagation of the 
energy-momentum-tensor, we use a dimension split approach. 
Therefore each work-item needs for its calculation only a one
dimensional neighborhood. The size of the 1-D differential stencil
still depends on the applied scheme, nevertheless this is a
significant reduction of the buffer
size\footnote{number of registers} needed. 
In this implementation each work-item computes independently
the geometric flux of all neighboring cells in order to calculate its
flux corrector and save the flux corrected value. In
figure~\ref{fig:neighbor} the needed neighborhood of each cell is
illustrated: to compute the flux corrected value for cell
$j$ (black) each work-item has to calculate the geometric propagated quantity
in cells $(j-2) \dots (j+2)$ (grey). Therefore additionally the cells $j-3$ and
$j+3$ have to be used.
This is done in the \texttt{for}-loop serially and buffered in the
\texttt{flux[]}-variable.
However, the loop increases the computational workload of each kernel
significantly, as each geometric flux has to be computed five times
more often than in the serial implementation. 
An additional overhead is caused by the calculation of both limited
anti-fluxes, i.e. $A_j$ and $A_{j-1}$ in equation~\eqref{eq:shasta-n}
in the variables \texttt{ea} and \texttt{eb} in listing~\ref{kernel},
hence this step doubles the workload compared to 
the serial implementation. Although the alternative
problem decomposition in section~\ref{alternative} avoids the
multiple calculation of geometric fluxes and anti-fluxes, its overall
performance proofed to be less due to the additional memory usage.
This is because the additional computations allows each work-item to
compute the propagated quantity independently from all other
work-items, and thus no serialization e.g. between the geometric
propagation and the anti-diffusion step is needed.

For each quantity a specialized kernel (similar to listing \ref{kernel})
is enqueued. To use the full capacity of a \textsmaller{GPU}, the
kernel size must be chosen carefully. If the kernel's active working set
is too big, only few work-items can run contemporary, as they must
share the available buffer memory. (Not only the neighborhood
(figure~\ref{fig:neighbor}) of the propagated quantity is necessary,
 but additionally each work-item needs to access a similar neighborhood
 of velocity, pressure, and different control variables. On the other
hand, if too many kernels are needed to compute the propagation, the
increased number of enqueued kernels propose an additional
orchestrating overhead. The use of mid weight kernels,
computing all needed fluxes of a neighborhood but propagating only one
quantity per kernel, allows for an efficient usage of the underlying
hardware (all stream cores in parallel) without overly increasing
the orchestration overhead. 

\subsubsection{Branching Free Execution Flow}
We have designed the components of \textsmaller{OpenCL-SHASTA} in a
modular way and implemented different kernels and auxiliary functions.
This allows shorter development cycles and a ensures good validation of the
algorithm. 
Additionally the  kernels and auxiliary functions can be
substituted even at run time leaving the choice of different anti-flux
functions, different source terms, and even different equations of
state. In (listing \ref{kernel}) the constant \texttt{diff} allows a
fine controlled application of numerical viscosity.
The value of the constant as well as the selection of more sophisticated or faster
anti-diffusion routines are controlled by meta-programming\cite{casi}.
As the kernels are compiled and loaded onto the \textsmaller{GPU} at run-time
this choices do not infer additional branchings, which would slow the
execution down. Nevertheless a free selection of the
desired numerical hydrodynamics realization is possible.
Therefore a wide variety of calculations can be carried out by the suggested
implementation without the need of complicated branching patterns
within the computational relevant parts of the program.

Let us stress the importance of avoiding (unnecessary) branching in
\textsmaller{GPU} programs: current \textsmaller{GPUs} do not offer a
\emph{branch predictor}, therefore branching comes generally with a significant penalty on \textsmaller{GPUs}. (As
stated in \cite{Daga:2011:AMO:2117686.2118481}
up to a loss of $30\%$ of the execution speed.)
To avoid branchings we have designe specialized kernels for all
quantities in \textsmaller{OpenCL-SHASTA}.
The kernels vary according to their propagated quantity, source terms
and propagation direction.
For example the kernels responsible for the momenta parallel to the propagation
direction have an additional source term, whereas momenta
perpendicular to the propagation direction are transported
source free. 
We use a number of halo cells to avoid complicated
indexing\footnote{Complicated indexing methods not only imply
  branchings but often a very inefficient memory access.}, for the
finite difference schemes at the border of the grid. The kernel operates only
within the boundaries of the inner grid, limited by grid size \textsmaller{(GS)} and halo size
\textsmaller{(HS)}. 
For stability reasons the code has been implemented based on a
half-step method. The different half steps are implemented without a
branching control structure, instead additional kernels change underlying
control variables like $\lambda$. We use a special double buffering
scheme to spare expensive memory copies. Instead of copying onto
different grids, we have designed to each kernel call an adjunct
kernel call, which reverses the used buffers. The adjunct
kernel calls are orchestrated in the launcher method (listing
\ref{enqueue}) of the \textsmaller{GPU}-class.

All different governing equations have been implemented in different
kernels in order to stay clear of any branches within the execution
flow. The orchestration of this set of 30 different kernel calls, is done by the
execution method on the host (listing \ref{enqueue}). Accordingly, the
\textsmaller{GPU} can start the computation, while the correct kernels
are still enqueued into the command que. No branching within the
\textsmaller{GPU} is necessary.

\subsubsection{Memory Aware 3-D Calculation}
Due to the dimension split scheme each operation is executed
only in one dimension at a time. This reduces the amount
of registers needed for each kernel drastically, as only 1-D
differential stencils (figure~\ref{fig:neighbor}) have to be
calculated. This reduces the needed memory size to a third compared to
a full 3-D stencil implementation.
The propagation direction follows a fixed permutation of the three axes.\footnote{We
  found this approach more stable than a Monte Carlo approach.}
After the initial copy of all needed quantities to its private memory,
the kernel (listing~\ref{kernel}) does not need any access to global 
memory for its calculations. On the contrary to the classical
\textsmaller{FORTRAN} implementation no differentials have to be
stored.  Intermediate results, like the geometric flux, anti-diffusion,
and flux-limiter can be stored in registers. The functions yielding
the necessary geometric factors (\texttt{qpt()} and \texttt{qmt()}) and
the antiflux (\texttt{antiflux()}) are inlined functions. They are
calculated exactly when needed (figure~\ref{fig:flowgraph}) and need
not to be calculated in advance.
(The former mentioned alternative problem decomposition (section~\ref{alternative}) needs
additional global memory for the geometric fluxes, anti-fluxes, and
flux limiters.)

According to the permutation scheme only the
propagation speed $v_\parallel$ parallel to the actual propagation direction
is calculated beforehand, which limits the global memory usage to only one field for the propagation
speed. This approach underlines again the \emph{recompute instead of memory lookup}
paradigm which holds for various applications on many-core
architectures\cite{kalcher}. The global memory footprint is
determined by the seven quantities residing
on a $~200^3$ cell grid. As only single precision is needed to
represent these quantities, the total memory consumption has been
reduced: including the double buffering memory scheme, the total
consumption is less than $500$ MB. This allows to run the
code even on commodity \textsmaller{GPUs}.
By limiting the working set to a minimum and concentrating on recompute
instead of expensive memory lookups, the remaining memory can be used
to increase the grid size, enhance precision or to hold more
complicated (tabled) equations of state, or even to enhance
\textsmaller{SHASTA} with the necessary tables to calculate viscous hydro
dynamics\cite{Molnar:2009tx,Niemi:2011ix}.

\section{Results}
We have tested different examples on the \textsmaller{LOEWE-CSC}
cluster using \textsmaller{AMD} 5870 \textsmaller{GPUs} and
\textsmaller{AMD} Opteron\textsuperscript{\texttrademark} 6172 processors (24 core).
Test cases included classic test problems, spheres of different metrics,  and
realistic initial conditions generated by \textsmaller{UrQMD}. For
realistic cases the physical simulation time is on the order of $10$ -- $20$ fm/c which
transforms to $200$ time steps (equalling to $16$ fm/c in the current
setup). 

The measured accelerations depend slightly on the geometry of the
input. The best acceleration is achieved for the realistic
\textsmaller{UrQMD} input files. Here the overall computing time for
a physical running time of $t=8$ fm/c for an Au+Au collision is reduced to less than 30 seconds.
Compared to the standard \textsmaller{FORTRAN}
implementation\cite{Petersen, Rischke} which
needs 1 hour and 15 minutes, we find an acceleration of more than a factor of 160.

\begin{figure}[]
\includegraphics[width=0.66\textwidth]{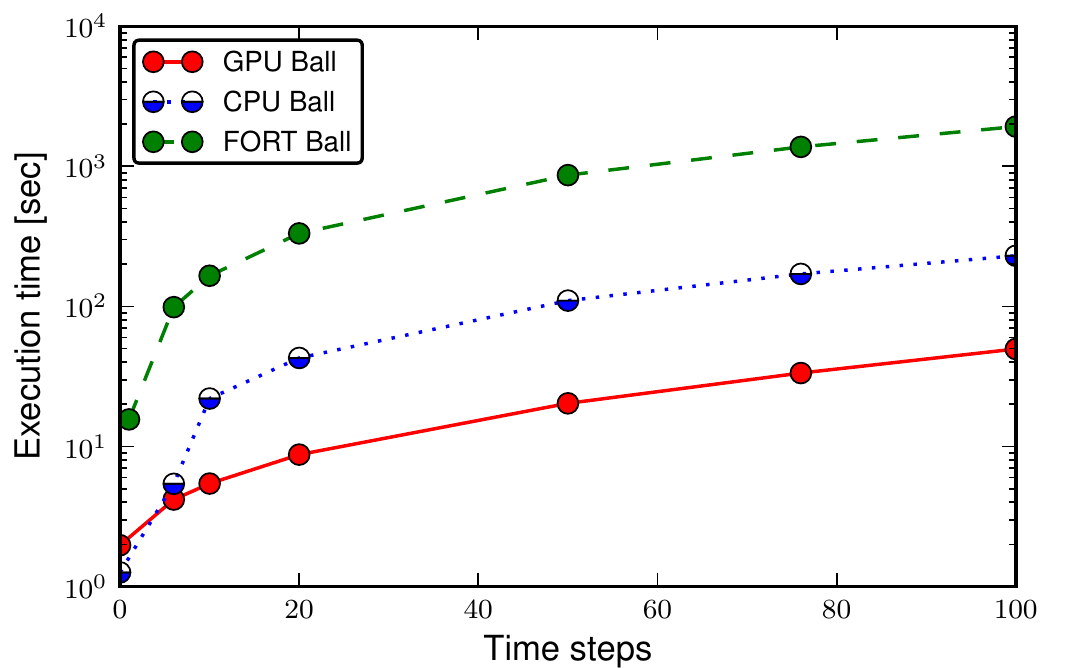}
\caption{\label{fig:total_time_ball} Total execution time for the expanding
  ball of $r=2$ fm with constant energy density. The \textsmaller{CPU} and
  \textsmaller{GPU} are measured with the exact same
  \textsmaller{OpenCL} code and compared to the Fortran
  \textsmaller{(FORT)} implementation\cite{Petersen, Rischke}. }
\end{figure}
\begin{figure}[]
\includegraphics[width=0.66\textwidth]{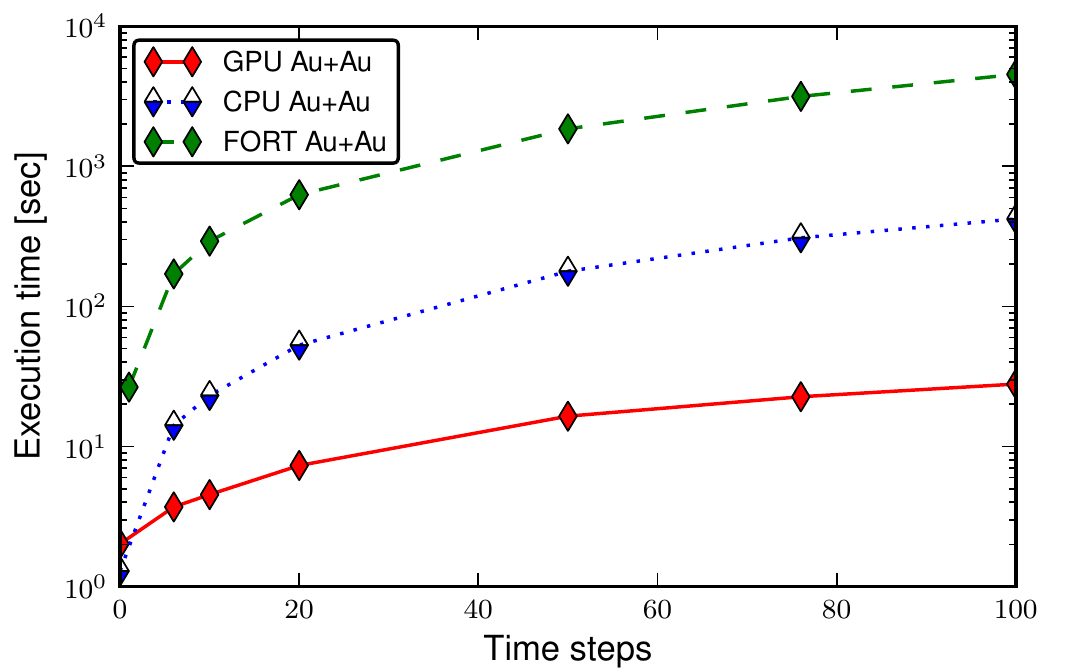}
\caption{\label{fig:total_time_pb} Total execution time for  a Au+Au
  collision with $\sqrt{\text{s}_\text{NN}}=200$ GeV.  The \textsmaller{CPU} and
  \textsmaller{GPU} are measured with the exact same
  \textsmaller{OpenCL} code and compared to the \textsmaller{FORTRAN}
  \textsmaller{(FORT)} implementation\cite{Petersen, Rischke}.}
\end{figure}
\begin{figure}[]
\includegraphics[width=0.666\textwidth]{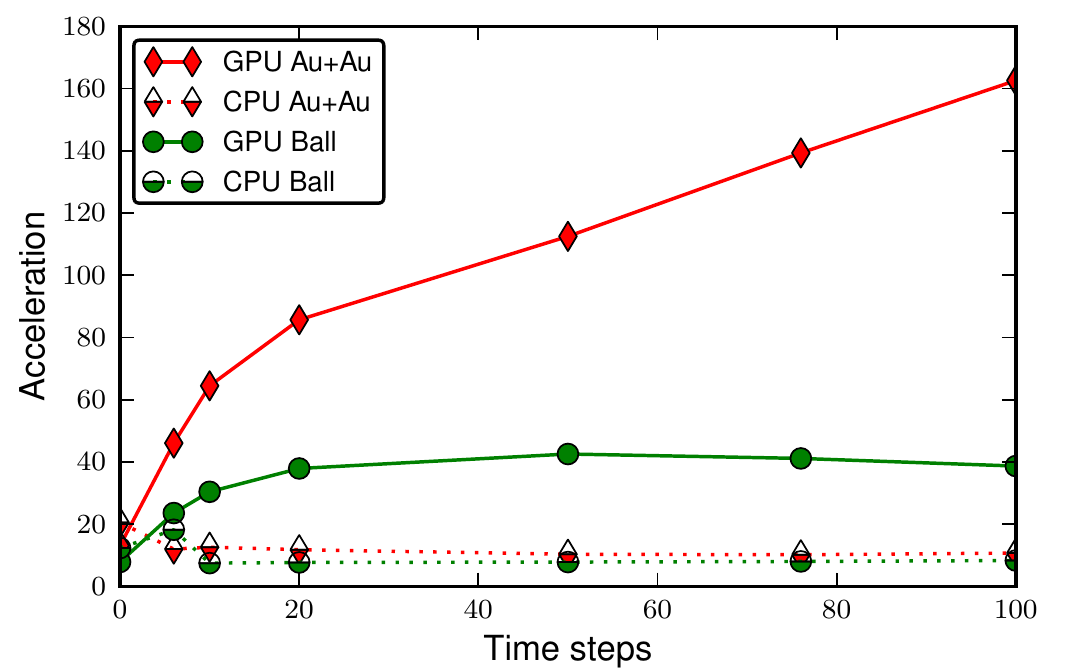}
\caption{\label{fig:acceleration} The acceleration of the execution
  speed for different initial geometries. Here again the exact same
  code is executed on \textsmaller{CPU} and \textsmaller{GPU}.}
\end{figure}
\begin{figure}
\includegraphics[width=0.666\textwidth]{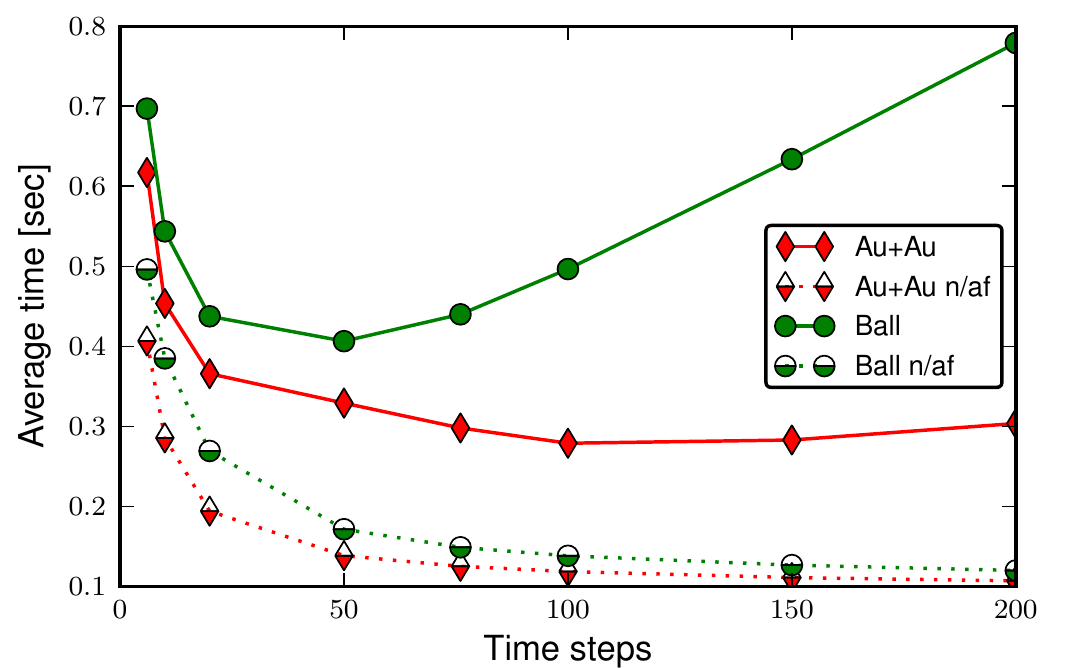}
\caption{\label{fig:antiflux} Average time consumption for one time step
  in the expansion of a spherical symmetric system (\emph{ball}) and for a Au+Au collision on
  the \textsmaller{GPU}. The \emph{n/af} routines show calculations without anti-flux.}
\end{figure}
In figure \ref{fig:total_time_ball} and figure \ref{fig:total_time_pb} the total execution time for two
different initial geometries is depicted.  Figure
\ref{fig:total_time_ball} shows the comparison for a spherical setup
(\emph{ball}, $\Vert \cdot \Vert_2$ ) of radius $2$ fm. Figure
\ref{fig:total_time_pb} shows the comparison between the
\textsmaller{FORTRAN} and \textsmaller{OpenCL} implementation on
\textsmaller{CPU/GPU} for the \textsmaller{UrQMD} initial
configuration for a Au+Au collision at $\sqrt{\text{s}_\text{NN}}=200$ GeV with impact parameter
$\text{b}=7$ fm.
Even without a \textsmaller{GPU} at hand the \textsmaller{OpenCL} implementation
provides a significant speed-up on \textsmaller{CPUs}. On the
\textsmaller{AMD} Opteron\textsuperscript{\texttrademark} 6172
processor (24 core), we measured an acceleration by a factor of
ten. Although this is not very efficient, let us stress that a simple
parallel execution of the standard \textsmaller{FORTRAN}
implementation is not possible, as the memory consumption of the
standard implementation is more than five times higher than the
\textsmaller{OpenCL} implementation.
Over the above the exact same implementation on \textsmaller{GPU} and
\textsmaller{CPU} are compared here. Though the program design allows
to choose appropriate kernels and environment variables for their
execution, depending on the provided architecture (see section~\ref{sec:problem}
and~\ref{sec:domain}), enabling further speed-ups on \textsmaller{CPUs}.

Figure \ref{fig:acceleration} summarizes the increasing speed-up of
the execution of the \textsmaller{OpenCL} implementation on the
\textsmaller{GPU}. The difference between the geometries is reflected
in the better distribution of computations to work-items. In the
spherical symmetric case (denoted as \emph{ball}) the initial geometry is concentrated
in the center of the grid, therefore at the beginning all the workload
is concentrated on few work-items. During the execution more cells are filled
with non-zero values. As each quantity in each cell is computed
separately in a work-item, a sparse grid is not very efficient, while a full grid
makes best use of all the stream cores of a \textsmaller{GPU}.

In figure \ref{fig:antiflux} we observe the initial cost of memory
transfer to the \textsmaller{GPU}, which becomes insignificant after
a small number of time steps. Nevertheless time consumption
increases again later, depending on the
problem's geometry. One observes the impact of the complex anti-flux
function on the average execution time. Without the complex anti-flux
no increase can be observed and the acceleration due filling the grid
takes fully place. In \textsmaller{SHASTA} the
anti-flux is corrected by a flux limiter. This flux limiter is
calculated by a search of maxima and minima of the surrounding cells
and fluxes towards this cells. In this calculation branching is
inherent and takes also place within different wavefronts. Therefore
all branches within these wavefronts are calculated by the device and the
correct results are gained by masking the wrong branches out. Hence
the execution time is increased significantly, when the flux limiter is not uniform. 

\begin{figure}
\includegraphics[width=0.666\textwidth]{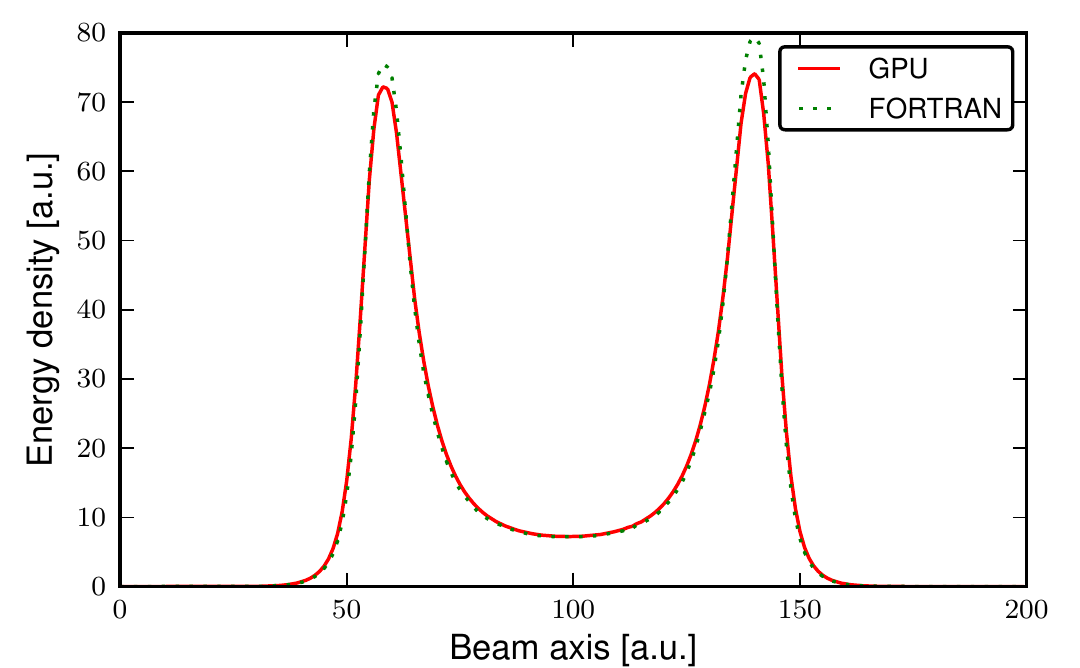}
\caption{\label{fig:numeric} Comparison between \textsmaller{FORTRAN}
  and \textsmaller{OpenCL} propagation for realistic initial
  conditions of a Au+Au collision with $\sqrt{\text{s}_\text{NN}}=200$ GeV and impact parameter
  $b=7$ fm at $t=8$ fm/c provided by \textsmaller{UrQMD}. (The asymmetry
    present in both implementations is due to fluctuations in the initial state.)}
\end{figure}

Finally figure \ref{fig:numeric}  shows the direct comparison between
the present \emph{single precision} implementation  (full line) and the
standard \textsmaller{FORTRAN} implementation (dotted line). For the
realistic initial setup of a $\sqrt{\text{s}_\text{NN}}=200$ GeV Au+Au collision provided by
\textsmaller{UrQMD} we find only minor differences between both implementations.
However, we work on a mixed precision implementation of
\textsmaller{OpenCL-SHASTA}. In this implementation we use
\emph{double precision} for the less stable parts of the numerics,
like the calculation of the boost. This becomes relevant at higher energies.
\section{Summary and Outlook}
On current hardware, like the \textsmaller{AMD} 5870
\textsmaller{GPU}, double precision  calculations come with a slowdown
of a factor of five\footnote{A special setup  is needed to use double
  precision as well as the increased memory consumption. These setups are
  highly hardware dependent and future hardware may be more suited to
  double precision calculations.}
, additionally a full double precision approach doubles the memory
consumption. We conclude: a single precision implementation allows
for fast calculations, as well as the enhancements of the underlying
physical model, e.g. by adding calculations of viscosity
\cite{Niemi:2011ix,Molnar:2009tx}.
If numerical instabilities occur, e.g. in the calculation
of the Lorentz-boost $\gamma$ we suggest the usage of 
mixed precision implementations.

We have designed the \textsmaller{OpenCL-SHASTA} to work on commodity
\textsmaller{GPUs}, nowadays present in almost every
computer\footnote{Even, where no such device is present, the
  implementation benefits of all present cores and vector units.}. 
As the \textsmaller{OpenCL} implementation allows for usage on
\textsmaller{GPUs}, accelerators, as well as on classical multi-core
processors, the \textsmaller{OpenCL-SHASTA} can be included in bigger
frameworks, that need to be executed on a variety of different architectures.  Due
to the tremendous speed-up it resolves the problem of a computationally demanding
hydrodynamic phase in hybrid models, like \textsmaller{UrQMD}, and allows for
better statistics, stability analyses\cite{stability}, and unprecedented event-by-event simulations.

\begin{acknowledgments}
The computational resources were provided by the \textsmaller{LOEWE-CSC}\cite{loewe}. This work
was supported by \textsmaller{HIC} for \textsmaller{FAIR} within the
Hessian \textsmaller{LOEWE} initiative.
\end{acknowledgments}
\bibliography{clpaper}
\pagebreak
\appendix
\section{Listings}
\lstset{language=C}
\begin{lstlisting}[caption={A generic kernel propagating source free
quantities in X-direction.}, label=kernel]
__kernel void generic_X( __global float* In, __global float* Out,
                         __global float* v, __global float* outcfl)
  uint idx = get_global_id(0);
  uint idy = get_global_id(1);
  uint idz = get_global_id(2);
  uint myid = idx + idy * DY + idz * DZ;
  if ( (idx >= HS) && (idy >= HS) && (idz >= HS) && (idx < GS - HS)
        && (idy < GS - HS) && (idz < GS - HS) ){		
    const float cfl = *outcfl ;
    const float diff = 1.0f;
    __private float basis[7] = {In[myid-3], In[myid-2], In[myid-1], 
                                In[myid], In[myid+1], In[myid+2], In[myid+3]};
    __private float flux[5];
    __private float velocity[7]= {v[myid-3] ,v[myid-2], v[myid-1],
                                  v[myid], v[myid+1], v[myid+2], v[myid+3]};
    for (short i = 0; i < 5; ++i){
      const float qpt = qp( velocity, i+1, cfl ); 
      const float qmt = qm( velocity, i+1, cfl );
      flux[i] = 0.5f * qpt*qpt * (basis[i+2]-basis[i+1]) 
              - 0.5f * qmt*qmt * (basis[i+1]-basis[i]) + (qpt+qmt)* basis[i+1];
    }
    const float ea = antiflux(flux, basis, 0, diff); 
    const float eb = antiflux(flux, basis, -1, diff);
    Out[myid] = flux[2] - ea + eb;
  } else Out[myid] = In[myid];
}
\end{lstlisting}

\lstset{language=C++}
\begin{lstlisting}[caption={A part of the enqueueing routine.}, label=enqueue]
 const int branch[] = {0,1,2, 0,2,1, 2,0,1, 2,1,0, 1,2,0, 1,0,2};
  for (int round = 0; round < steps; ++round) {
    i = branch[round % 18];
    if ( i == 0 ) { 
      queue.enqueueNDRangeKernel(untangT, cl::NullRange,
                                 cl::NDRange(GS,GS,GS), cl::NullRange,
                                 NULL, NULL);
      queue.enqueueNDRangeKernel(half_cfl, cl::NullRange,
                                 cl::NDRange(1,0,0), cl::NullRange,
                                 NULL, NULL);
      queue.enqueueNDRangeKernel(propX_e, cl::NullRange,
                                 cl::NDRange(GS,GS,GS), cl::NullRange,
                                 NULL, NULL);
      queue.enqueueNDRangeKernel(prop_parallel_mx, cl::NullRange,
                                 cl::NDRange(GS,GS,GS), cl::NullRange,
                                 NULL, NULL);
      queue.enqueueNDRangeKernel(propX_my, cl::NullRange,
                                 cl::NDRange(GS,GS,GS), cl::NullRange,
                                 NULL, NULL);
  }
\end{lstlisting}
\pagebreak
\section{Parallelization Guidelines}
Of course no simple and general mechanism of
parallelizing algorithms can be stated. Nevertheless, most of the
steps we investigated on for the \textit{OpenCL-SHASTA} can be applied to
similar algorithms in this field. Therefore we state here a \emph{best
  practice} approach, we believe can be followed by other research
groups, who wish to port their code on \textsmaller{GPUs}.
\subsection{Problem Decomposition}
\begin{itemize}
\item Physical quantities, that by superposition are not dependent on
  each other, are the first choice to be computed in parallel with different
  kernels. (These are e.g. $\mathcal{E}$, $\boldsymbol{\mathcal{M}}$, and
  $\mathcal{N}$.)
\item If the independence between variables can be gained by a finer
  time step resolution, the additional execution time of more time
  steps is easily compensated by the parallelization.
\item Not only physical quantities can be parallelized. Also
  intermediate results within a numerical scheme can often be
  calculated in parallel.
\item A data dependence graph, like figure~\ref{fig:flowgraph}, helps to
group calculations to kernels. (Often the connectivity between cliques
is a first hint to beneficial cuts)
\end{itemize}
\subsection{Domain Decomposition}
\begin{itemize}
\item In grid based algorithms a decomposition to work-items for each grid
point is a natural choice. (We group  $206 \times 206 \times 206$
work-units per kernel to the queue, for the $200^3$-cell grid, adding
halo cells.)
\item The aim of the decomposition is to minimize
the communication between work-items: numerical schemes,
that need a high communication between grid cells might be more
efficient with work-items handling small neighborhoods of the grid.
(In \textsmaller{OpenCL-SHASTA's} kernels the initial step is to store
\textbf{all} necessary quantities in local memory, thus no further
communication to other cells is necessary.)
\item  In \textsmaller{OpenCL} exists the possibility to group kernels to
workgroups, simplifying and accelerating communication within
work-items of workgroup.
\item Here also a data dependence graph for the spacial dependencies
  can show how a useful grouping can be done.
\item In algorithms not based on a grid, but e.g. test-particles a
  possible choice is to map the number of particles to work-items.
\item The ordering of the particles should mimic the spacial
ordering (in phase-space) of the particles, to benefit from potential
cache and coalescent memory lookup effects.
\end{itemize}
\subsection{Kernel Design Criteria}
\begin{itemize}
\item Depending on the number of work-items that shall be executed in
  parallel, the active working set of a kernel may not be too big, as
  the ressources are divided by all contemporary executed kernels. (We
  chose to compute only \textbf{one} physical quantity per kernel, as
  many fluxes have to be computed. Algorithms with a smaller update
  step maybe more efficient e.g. by computing the momenta together.)
\item Kernels should be as independent as possible, synchronization
  between virtually thousands of work-items is cumbersome and error
  prone. (We have chosen to calculate all fluxes needed within each
  kernel, thus no communication was necessary.)
\item A favorable approach is to start with smallest possible kernels
  and merging kernels whilst measuring the execution time. (For example in the
\textsmaller{OpenCL-SHASTA} we started with extra kernels calculating
the geometric flux and kernels calculating the anti-flux. The merge of
both kernels proofed to be more efficient.)
\end{itemize}
\subsection{Hardware Specific Optimizations}
\begin{itemize}
\item Branching within kernels is most detrimental on
  \textsmaller{GPUs}, codes designed for \textsmaller{CPUs} are less
  affected by branchings. 
\item Often the needed branchings in physical models follow a fixed
  scheme completely determined before execution, e.g. the model's choice
  of the EoS. Therefore it should be encoded with different kernels and
  orchestrated in advance. (The kernels for velocity and pressure can
  easily be substituted before they are loaded to the \textsmaller{GPU}.) 
\item Memory access is always a critical point: it should be avoided
  wherever possible. (Many simulation codes save differentials in
  arrays, they can be calculated directly)
\item Floating point calculations are highly efficient on
  \textsmaller{GPUs}. Even transcendental calculations should not be
  stored in arrays with intermediate results.
\item Instead the use of inlined functions (like
e.g. \texttt{qp} and \texttt{qm} in \textsmaller{OpenCL-SHASTA})
allows for a clean code and an efficient execution.
\item Caching might be an issue on special hardware, spacial locality
  ensures often an efficient usage of present cache structures.
\item Coalesced memory access is also often gained by a correctly
  ordered enqueueing. (Spacial neighbors should have neighboring
  work-item-ids.)
\end{itemize}
\end{document}